\documentclass{emulateapj}
\shorttitle{Early Afterglow of GRB 050801}
\shortauthors{Rykoff et al.}

\begin{document}

\title{The Anomalous Early Afterglow of GRB 050801}

\author{
E.~S.~Rykoff,\altaffilmark{1},
V.~Mangano,\altaffilmark{2},
S.~A.~Yost\altaffilmark{1},
R.~Sari\altaffilmark{3},
F.~Aharonian\altaffilmark{4},
C.~W.~Akerlof\altaffilmark{1},
M.~C.~B.~Ashley\altaffilmark{5},
S.~D.~Barthelmy\altaffilmark{6},
D.~N.~Burrows\altaffilmark{7},
N.~Gehrels\altaffilmark{6},
E.~G\"{o}\v{g}\"{u}\c{s}\altaffilmark{8},
D.~Horns\altaffilmark{4},
\"{U}.~K{\i}z{\i}lo\v{g}lu\altaffilmark{10},
H.~A.~Krimm\altaffilmark{6,11},
T.~A.~McKay\altaffilmark{1},
M.~\"{O}zel\altaffilmark{12},
A.~Phillips\altaffilmark{5},
R.~M.~Quimby\altaffilmark{13},
G.~Rowell\altaffilmark{4},
W.~Rujopakarn\altaffilmark{1},
B.~E.~Schaefer\altaffilmark{14},
D.~A.~Smith\altaffilmark{15},
H.~F.~Swan\altaffilmark{1},
W.~T.~Vestrand\altaffilmark{16},
J.~C.~Wheeler\altaffilmark{13},
J.~Wren\altaffilmark{16},
F.~Yuan\altaffilmark{1},
}

\altaffiltext{1}{University of Michigan, 2477 Randall Laboratory, 450 Church
        St., Ann Arbor, MI, 48109, erykoff@umich.edu}
\altaffiltext{2}{INAF-IASF, Palermo, Italy}  
\altaffiltext{3}{California Institute of Technology, Pasadena, CA, 91125, USA}
\altaffiltext{4}{Max-Planck-Institut f\"{u}r Kernphysik, Saupfercheckweg 1,
        69117 Heidelberg, Germany}
\altaffiltext{5}{School of Physics, Department of Astrophysics and Optics,
        University of New South Wales, Sydney, NSW 2052, Australia}
\altaffiltext{6}{NASA Goddard Space Flight Center, Laboratory for High Energy
        Astrophysics, Greenbelt, MD 20771}
\altaffiltext{7}{Pennsylvania State University, University Park, PA, 16802,
        USA}
\altaffiltext{8}{Sabanc{\i} University, Istanbul, Turkey}
\altaffiltext{9}{Istanbul University Science Faculty, Department of Astronomy
        and Space Sciences, 34119, University-Istanbul, Turkey}
\altaffiltext{10}{Middle East Technical University, 06531 Ankara, Turkey}
\altaffiltext{11}{Universities Space Research Association, 10227 Wincopin
        Circle, Suite 212, Columbia, MD 21044}
\altaffiltext{12}{\c{C}anakkale Onsekiz Mart \"{U}niversitesi, Terzio\v{g}lu
        17020, \c{C}anakkale, Turkey}
\altaffiltext{13}{Department of Astronomy, University of Texas, Austin, TX
        78712}
\altaffiltext{14}{Department of Physics and Astronomy, Louisiana State
        University, Baton Rouge, LA 70803}
\altaffiltext{15}{Guilford College, Greensboro, NC, 27410, USA}
\altaffiltext{16}{Los Alamos National Laboratory, NIS-2 MS D436, Los Alamos, NM
        87545}

\begin{abstract}

The ROTSE-IIIc telescope at the H.E.S.S. site, Namibia, obtained the earliest
detection of optical emission from a Gamma-Ray Burst (GRB), beginning only
21.8~s from the onset of \emph{Swift} GRB~050801.  The optical lightcurve does
not fade or brighten significantly over the first $\sim250$ s, after which
there is an achromatic break and the lightcurve declines in typical power-law
fashion.  The \emph{Swift}/XRT also obtained early observations starting at
69~s after the burst onset.  The X-ray lightcurve shows the same features as
the optical lightcurve.  These correlated variations in the early optical and
X-ray emission imply a common origin in space and time.  This behavior is
difficult to reconcile with the standard models of early afterglow emission.

\end{abstract}
\keywords{gamma rays:bursts}

\section{Introduction}

Gamma-ray bursts (GRBs) are the most luminous explosions in the universe, but
the origin of their emission remains elusive.  With the launch of the
\emph{Swift} $\gamma$-ray Burst Explorer~\citep{gcgmn04} in late 2004, great
progress has been made in the study of the early afterglow phase of GRBs.
However, only a small number of bursts have been imaged simultaneously in both
the optical and X-ray bands in the first minutes after the
burst~\citep{nkgpg05,qryaa05,rykaa05,bbbbc05}.

In this letter, we report on the earliest detection of optical emission,
starting at 21.8 seconds after the onset of GRB~050801 with the ROTSE-IIIc
(Robotic Optical Transient Search Experiment) telescope located at the
H.E.S.S. site in Namibia.  This is the most densely sampled early lightcurve
yet obtained.  It does not fade or brighten significantly over the first
$\sim250$ seconds, after which there is a break and the lightcurve declines in
a typical power-law fashion.  The \emph{Swift}/XRT also obtained early
observations starting at 69 seconds after the burst onset.  The X-ray
lightcurve shows the same features as the optical lightcurve.  These correlated
variations in the early optical and X-ray emission imply a common origin in
space and time.  This behavior differs from that seen in
GRB~050319~\citep{qryaa05}, GRB~050401~\citep{rykaa05}, and
GRB~050525a~\citep{bbbbc05}.  It is difficult to explain this behavior with
standard models of early afterglow emission without assuming there is
continuous late time injection of energy into the afterglow.

\section{Observations and Analysis}
\label{sec:observations}

The ROTSE-III array is a worldwide network of 0.45~m robotic, automated
telescopes, built for fast ($\sim 6$ s) responses to GRB triggers from
satellites such as HETE-2 and \emph{Swift}.  They have wide ($1\fdg85 \times
1\fdg85$) fields of view imaged onto Marconi $2048\times2048$ back-illuminated
thinned CCDs, and operate without filters.  The ROTSE-III systems are described
in detail in \citet{akmrs03}.

On 2005 August 01, \emph{Swift}/BAT detected GRB~050801 (\emph{Swift} trigger
148522) at 18:28:02.1 UT.  The position was distributed as a Gamma-ray Burst
Coordinates Network (GCN) notice at 18:28:16 UT, with a $4\arcmin$ radius
$3\sigma$ error circle.  The burst had a $T_{90}$ duration of
$20\pm3\,\mathrm{s}$ in the 15-350 keV band, and consisted of two peaks
separated by around 3 seconds.  The position was released during the tail end
of the $\gamma$-ray emission~\citep{smbbc05}.  The \emph{Swift} satellite
immediately slewed to the target, with the XRT beginning observations in
windowed timing mode at 69 s after the start of the burst and switching to
photon counting mode at 89.3 s after the trigger.

ROTSE-IIIc, at the H.E.S.S. site in Namibia, responded automatically to the GCN
notice, beginning its first exposure in less than 8 s, at 18:28:23.9 UT. The
automated burst response included a set of ten 5-s exposures, ten 20-s
exposures, and 134 60-s exposures before the burst position dropped below our
elevation limit.  The first set of ten exposures were taken with subframe
readout mode to allow rapid sampling (3-s readout between each 5-s exposure).
Near real-time analysis of the ROTSE-III images detected a $15^{th}$ magnitude
source at $\alpha=13^h36^m35\fs4$, $\delta=-21\arcdeg55\arcmin42\farcs0$
(J2000.0) that was not visible on the Digitized Sky Survey red plates, which we
reported via the GCN Circular e-mail exploder within 7 minutes of the
burst~\citep{ryr05}. No spectroscopic redshift has been reported for
this GRB, although the \emph{Swift}/UVOT detected the afterglow in all filters
including the $UVW2$ filter at $188\,\mathrm{nm}$~\citep{bbhgc05}, which implies
that the redshift is $\lesssim1.2$.  In addition, the afterglow was dimmer than
23 mag with no evidence for a bright host galaxy~\citep{fjhww05b}.

The X-ray photometry is shown in Table~\ref{tab:xray}.  Time bin midpoints and
durations are listed in seconds, relative to the \emph{Swift} trigger time,
18:28:02 UT.  The count rate is in counts/s and the flux is in
$10^{-11}\,\mathrm{erg}\,\mathrm{cm}^{-2}\,\mathrm{s}^{-1}$, for the energy
range 0.2-10 keV.  The X-ray data has been corrected for a hot CCD column
crossing the source as well as a nearby source $30''$ away.  Photon counting
data from the first orbit have been corrected for pile-up.  We chose a time
binning that ensures a detection of at least $3.5\sigma$ for each time bin
before corrections were applied.  The gaps in the data are caused by earth
occultation.  There is no spectral variation across the lightcurve, and the
$N_H$ value is consistent with the Galactic value
($7\times10^{20}\,\mathrm{cm}^{-2}$).  The best-fit spectrum (with $N_H$ fixed
to $7\times10^{20}\,\mathrm{cm}^{-2}$) is a power law with photon index
$1.87\pm0.15$ (90\% confidence level).  The relative errors for the fluxes are
slightly larger than those for the count rate due to the additional systematic
error from the conversion.

\begin{deluxetable}{cccc}
\tablewidth{0pt}
\tablecaption{\emph{Swift}/XRT observations of the afterglow
  of GRB~050801.\label{tab:xray}}
\tablehead{
\colhead{T-mid (s)} & \colhead{Duration (s)} & \colhead{Count Rate
  ($\mathrm{cts}\,\mathrm{s}^{-1}$)} & 
\colhead{Flux ($10^{-11}\,\mathrm{erg}\,\mathrm{cm}^{-2}\,\mathrm{s}^{-1}$)}
}
\startdata
74.1 &     10.0 & $3.46\pm0.81$ & $19.2\pm 5.7$ \\
84.1 &     10.0 & $1.99\pm0.70$ & $11.1\pm 4.4$ \\
111.8 &     45.0 & $1.62\pm0.39$ & $ 9.0\pm 2.7$ \\
164.3 &     60.0 & $1.40\pm0.31$ & $ 7.8\pm 2.2$ \\
214.3 &     40.0 & $1.83\pm0.43$ & $10.1\pm 3.0$ \\
254.3 &     40.0 & $1.83\pm0.43$ & $10.1\pm 3.0$ \\
291.8 &     35.0 & $2.25\pm0.51$ & $12.5\pm 3.6$ \\
334.3 &     50.0 & $1.52\pm0.35$ & $ 8.4\pm 2.5$ \\
396.8 &     75.0 & $1.01\pm0.24$ & $ 5.6\pm 1.6$ \\
471.8 &     75.0 & $1.05\pm0.24$ & $ 5.8\pm 1.7$ \\
561.8 &    105.0 & $0.69\pm0.167$ & $ 3.8\pm 1.1$ \\
686.8 &    145.0 & $0.50\pm0.12$ & $ 2.76\pm 0.83$ \\
859.3 &    200.0 & $0.31\pm0.08$ & $ 1.71\pm 0.55$ \\
4346.7 &    320.0 & $0.069\pm0.021$ & $ 0.38\pm 0.14$ \\
4856.7 &    700.0 & $0.060\pm0.013$ & $ 0.333\pm 0.092$ \\
5715.3 &    510.0 & $0.043\pm0.014$ & $ 0.241\pm 0.086$ \\
6357.8 &    775.0 & $0.027\pm0.009$ & $ 0.149\pm 0.055$ \\
11249.6 &   2560.0 & $0.012\pm0.003$ & $ 0.065\pm 0.021$ \\
17040.6 &   2550.0 & $0.0095\pm0.0028$ & $ 0.053\pm 0.018$ \\
22814.0 &   2575.0 & $0.0069\pm0.0025$ & $ 0.038\pm 0.015$ \\
31556.0 &   8237.1 & $0.0049\pm0.0015$ & $ 0.0275\pm 0.0096$ \\
47812.8 &  17585.3 & $0.0033\pm0.0010$ & $ 0.0182\pm 0.0063$ \\
366425.9 & 515983.6 & $< 0.0004$ & $< 0.00222$ \\

\enddata
\tablecomments{Time bin midpoints and durations are relative
  to the \emph{Swift} trigger time, 18:28:02 UT.}
\end{deluxetable}

The optical photometry is shown in Table~\ref{tab:opt}.  The ROTSE-IIIa images
were bias-subtracted and flat-fielded by our automated pipeline.  The
flat-field image was generated from 30 twilight images.  We used
SExtractor~\citep{ba96} to perform the initial object detection and to
determine the centroid positions of the stars.  The images were processed with
our custom RPHOT photometry program based on the DAOPHOT PSF-fitting photometry
package~\citep{qryaa05}. The unfiltered thinned ROTSE-III CCDs have a peak
response similar to an $R$-band filter.  The magnitude zero-point for was
calculated from the median offset of the fiducial reference stars to the USNO
B1.0 $R$-band measurements to produce $C_R$ magnitudes. After the first 30
images, frames were co-added in logarithmic time bins to maintain roughly
constant signal-to-noise.

\begin{deluxetable}{ccc}
\tablewidth{0pt}
\tablecaption{ROTSE-IIIc Optical Photometry of GRB~050801.\label{tab:opt}}
\tablehead{
\colhead{$t_{\mathrm{start}}$} &
\colhead{$t_{\mathrm{end}}$} &
\colhead{$C_R$}
}
\startdata
          21.8 &         26.8 & $14.93\pm 0.05$\\
          29.9 &         34.9 & $14.79\pm 0.05$\\
          38.0 &         43.0 & $14.80\pm 0.04$\\
          46.1 &         51.1 & $14.91\pm 0.06$\\
          54.2 &         59.2 & $14.83\pm 0.05$\\
          62.4 &         67.4 & $14.91\pm 0.04$\\
          70.5 &         75.5 & $14.75\pm 0.04$\\
          78.6 &         83.6 & $14.87\pm 0.05$\\
          86.7 &         91.7 & $14.88\pm 0.05$\\
          94.8 &         99.8 & $14.93\pm 0.05$\\
         113.5 &        133.5 & $14.98\pm 0.03$\\
         143.3 &        163.3 & $15.09\pm 0.03$\\
         172.7 &        192.7 & $15.12\pm 0.03$\\
         203.0 &        223.0 & $15.06\pm 0.03$\\
         232.5 &        252.5 & $15.13\pm 0.04$\\
         262.3 &        282.3 & $15.21\pm 0.04$\\
         291.8 &        311.8 & $15.35\pm 0.04$\\
         321.0 &        341.0 & $15.47\pm 0.04$\\
         350.8 &        370.8 & $15.59\pm 0.03$\\
         380.3 &        400.3 & $15.70\pm 0.04$\\
         409.9 &        469.9 & $15.89\pm 0.04$\\
         479.8 &        539.8 & $16.12\pm 0.03$\\
         549.0 &        609.0 & $16.29\pm 0.04$\\
         618.2 &        678.2 & $16.31\pm 0.05$\\
         688.1 &        748.1 & $16.63\pm 0.06$\\
         757.2 &        817.2 & $16.59\pm 0.06$\\
         826.6 &        886.6 & $16.66\pm 0.07$\\
         896.3 &        956.3 & $16.75\pm 0.06$\\
         965.5 &       1025.5 & $16.93\pm 0.07$\\
        1034.9 &       1094.9 & $16.92\pm 0.09$\\
        1104.7 &       1233.9 & $16.99\pm 0.06$\\
        1243.6 &       1442.0 & $17.10\pm 0.05$\\
        1451.4 &       1650.3 & $17.39\pm 0.07$\\
        1659.7 &       1858.6 & $17.48\pm 0.07$\\
        1867.9 &       2136.8 & $17.60\pm 0.06$\\
        2146.5 &       2485.3 & $17.78\pm 0.07$\\
        2495.2 &       2832.6 & $17.88\pm 0.07$\\
        2841.9 &       3249.7 & $18.26\pm 0.11$\\
        3259.7 &       3736.8 & $18.24\pm 0.09$\\
        3745.9 &       4332.1 & $18.71\pm 0.20$\\
        4341.4 &       4956.6 & $18.49\pm 0.09$\\
        4966.5 &       5721.7 & $18.88\pm 0.12$\\
        5731.0 &       6554.7 & $18.99\pm 0.15$\\
        6564.4 &       7527.4 & $18.83\pm 0.13$\\
        7536.7 &       8619.8 & $19.63\pm 0.22$\\
        8629.6 &      10357.0 & $19.49\pm 0.16$\\
\enddata
\tablecomments{Start and end times are relative to the \emph{Swift} trigger
  time, 18:28:02 UT.}
\end{deluxetable}

\section{Results}

With a detection only 21.8 s after the start of the burst, this is the earliest
detection of an optical counterpart of a GRB, as well as the most densely
sampled early afterglow.  Only four GRBs have had optical counterparts detected
within the first minute, and none of these had more than two detections in the
first minute.  The first 250 s of the optical afterglow shows short timescale
variability relative to an overall flat lightcurve.  This is in stark contrast
to the prompt counterpart of GRB~990123~\citep{abbbb99}, which had a very bright
$9^{th}$ mag peak at 60~s after the burst onset, generally interpreted as the
signature of reverse shock emission~\citep{sp99b}.  This afterglow shows no
evidence for reverse shock emission.  

\begin{figure*}
\rotatebox{90}{\scalebox{0.85}{\plotone{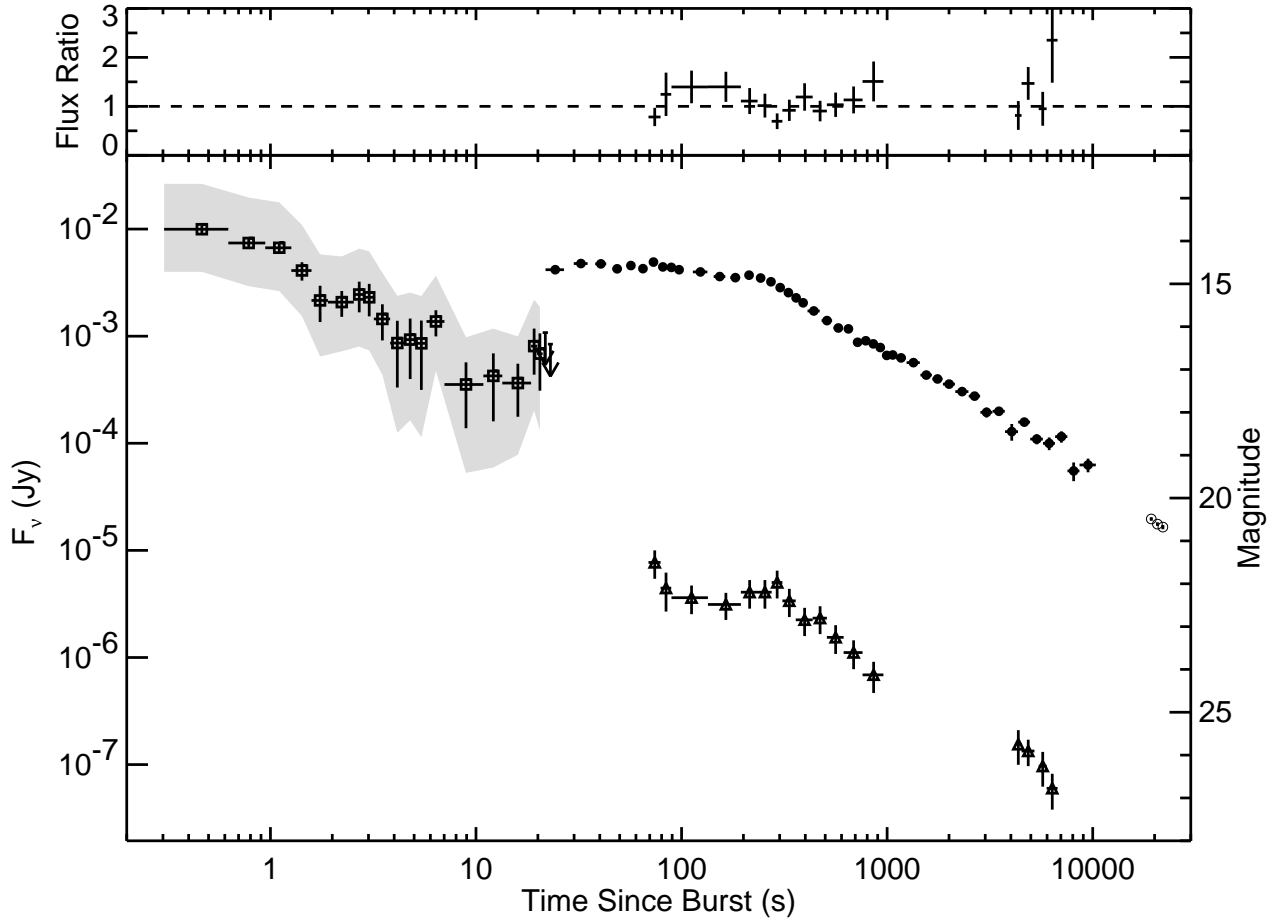}}}
\caption{\label{fig:optandxraylc}Comparison of the early optical and X-ray
  lightcurves of GRB~050801.  The main panel shows the optical and X-ray
  lightcurves.  The X-ray flux densities (triangles) are supplemented by the
  prompt BAT $\gamma$-ray flux densities (squares) extrapolated to the X-ray
  band [0.2-10 keV].  The ROTSE-III optical magnitudes (solid circles) have
  been converted to flux densities assuming the unfiltered ROTSE-III images are
  equivalent to $R_c$.  Where the error bars are not visible they are smaller
  than the plot symbols.  The empty circles are from the Danish 1.5~m
  telescope~\citep{fjhww05}, and are consistent with the ROTSE-III decay slope
  after 1500~s.  The magnitude scale is on the right for reference.  The top
  panel shows the ratio of optical flux to X-ray count rate for the first 7000
  s, scaled to the average ratio value.  The X-ray count rate rather than the
  X-ray flux was used to avoid the systematic error introduced when converting
  from count rate to flux, and is made possible by the lack of X-ray spectral
  evolution.  The ROTSE-III observations have been co-added to match the times
  of the XRT integrations as closely as possible.}
\end{figure*}

Figure~\ref{fig:optandxraylc} shows a comparison of the early optical and X-ray
lightcurves of GRB~050801, combined with the prompt $\gamma$-ray emission.  The
prompt BAT $\gamma$-ray flux densities have been extrapolated to the X-ray band
[0.2-10 keV].  This extrapolation was performed with the best-fit photon index
of $2.0\pm0.2$ for the time-averaged $\gamma$-ray spectrum from 20-150 keV, as
in \citet{tgcmc05}.  The statistical errors scaled from the BAT count rate are
shown; the gray region denotes the uncertainties from the extrapolation to the
X-ray regime.  The X-ray flux values have been converted to flux density (Jy)
using an effective frequency of $<\nu> = 6.89\times10^{17}\,\mathrm{Hz}$, the
flux weighted average in the 0.2-10 keV range with the best-fit photon index
$\Gamma = 1.87$.  The ROTSE-III optical magnitudes have been converted to flux
density assuming the unfiltered ROTSE-III images are equivalent to $R_c$, and
have been approximately adjusted for Galactic extinction by 0.24
mag~\citep{sfd98}. The de-extinction does not have a significant effect on the
derived spectral indices.  After the break at $\sim250$ seconds, the optical
lightcurve decays as $t^{-1.31\pm0.11}$, followed by a brief but significant
plateau at $\sim800$ seconds.  The top panel shows the ratio of optical flux to
X-ray count rate for the first 7000 s, scaled to the average ratio value.  The
X-ray count rate rather than the X-ray flux was used to avoid the systematic
error introduced when converting from count rate to flux, and is made possible
by the lack of X-ray spectral evolution.  The ROTSE-III observations have been
co-added to match the times of the XRT integrations as closely as possible.
The flux ratio is consistent with a constant value (dashed-line) with a
$\chi^2$ of 15.9 (16 degrees of freedom).  The break at $\sim250\,\mathrm{s}$
has no systematic change in the optical to X-ray flux ratio, and is therefore
achromatic.

Only three $\gamma$-ray bursts have had prompt optical detections
contemporaneous with the $\gamma$-ray emission.  The prompt optical counterpart
of GRB~041219a~\citep{vwwfs05} was correlated with the $\gamma$-ray emission,
implying a common origin.  However, both GRB~990123~\citep{abbbb99} and
GRB~050401~\citep{rykaa05} demonstrated a different origin for the $\gamma$-rays
and the optical radiation.  Although we do not have a prompt optical detection
in the case of GRB~050801, we can interpolate between the high energy prompt
lightcurve scaled to the X-ray band (gray band in Figure~\ref{fig:optandxraylc})
and the first X-ray detection.  During this interval the high energy emission
falls by a factor of $\gtrsim100$ while the optical emission is unchanged.
This suggests a different origin for the prompt $\gamma$-ray emission and the
early optical emission.  However, the X-ray and optical afterglow of GRB~050801
do appear to arise from a similar origin after $\sim80$~s.  The two lightcurves
are plotted in the main panel of Figure~\ref{fig:optandxraylc}.  Each lightcurve
shows similar flat behavior at the early time, with a break around 250~s.

\section{Discussion}

In the standard fireball model of GRB afterglow emission, the spectral energy
distribution of GRB afterglows can be fit by a broken power-law with spectral
segments $F_\nu \propto \nu^\beta$ (for a review, see \citet{p05}).  The
spectral index obtained by comparing the de-extincted optical (see
Figure~\ref{fig:optandxraylc}) to X-ray flux density during the second XRT
integration is $\beta_{\mathrm{opt-X}} = -0.92\pm0.05$, consistent with the
X-ray only spectral index of $\beta_{\mathrm{X}} = -0.87\pm0.15$ [0.2-10 keV].
To test for evolution in the broadband spectral index, we have compared the
optical and X-ray lightcurves during the first 7000 s (top panel of
Figure~\ref{fig:optandxraylc}).  The optical to X-ray flux ratio is consistent
with a constant value ($\chi^2=15.9$ with 16 degrees of freedom).  Across the
break at 250 s, both $\beta_{\mathrm{opt-X}}$ and $\beta_{\mathrm{X}}$ are
unchanged, and therefore the break is achromatic.  Furthermore, there is no
evidence of a spectral change in the UVOT images~\citep{bbhgc05}, although the
time resolution is insufficient to constrain the time of the break.  Many X-ray
lightcurves have been seen to steepen around 1000 s - 5000 s post-burst with no
change in the X-ray spectral index~\citep{nkgpg05}, For the few bursts with
sufficient early optical and X-ray coverage~\citep{qryaa05, bbbbc05}, this
behavior has not been mirrored in the optical band.

The tight correlation between the optical and X-ray emission suggests that they
share the same origin in space and time.  The standard fireball model of GRB
afterglows can explain the behavior of the optical and X-ray lightcurve after
250~s.  The observed spectral parameters and decay indices are most consistent
with a fireball expanding adiabatically into a constant density medium, with
the typical synchrotron frequency $\nu_\mathrm{m}$ below the optical band, and
the cooling frequency $\nu_\mathrm{c}$ above the X-ray band. For example, this
can be produced by the following parameters: the electron energy index $p =
2.8$; the isotropic equivalent energy $E \sim 10^{53}\,\mathrm{erg}$ at a
redshift of $z \sim 0.5$; the circumburst density $n \sim
0.7\,\mathrm{cm}^{-3}$; the energy fraction in the electrons $\epsilon_e \sim
0.07$; and the energy fraction in the magnetic field $\epsilon_B \sim
0.0002$. These values of the electrons and magnetic energy are consistent with
those deduced for other bursts albeit on the lower side.  If the ejecta were
expanding into a $1/r^2$ density profile (a so-called ``wind'' medium), the
fireball model predicts a relationship between the spectral and temporal
behavior that is inconsistent at the $4\sigma$ level with the observations
after 250~s.

We now investigate the possible explanations of the flat early lightcurve and
the origin of the break at 250~s.  First, any spectral transition (e.g. $\nu_m$
crossing the optical band) would fail to explain the achromatic nature of the
break. Achromatic breaks observed in other afterglows have been interpreted as
geometric, when the edge of a conical jet becomes visible to the observer and
the jet starts to spread~\citep{hbfsk99,sgkpt99}. At 250 s, this would be the
earliest such ``jet break'' detected.  In the fireball model, the post jet
break afterglow is expected to decay as $t^{-p}$, where $p$ is the electron
energy index with $N_e \propto E^{-p}$, provided that $p>2$~\citep{sph99}. A
hard electron index of $p<2$ predicts a post-jet decay even steeper than
$t^{-p}$~\citep{dc01}.  Therefore, the observed post-break temporal decay
implies $p \le 1.3$ which predicts a significant pre-break decay~\citep{dc01}
that is inconsistent with the observed pre-break flatness as well as the
observed spectral index $\beta$.  Therefore, the achromatic evolution of
GRB~050801 cannot be explained with a jet break.

We have investigated whether the early afterglow is consistent with the
predictions of a structured jet viewed off axis~\citep{gk03}.  In this case, it
is difficult to create a sharp early break; under such conditions, the
post-break evolution should track closely with the electron energy index $p$,
which is inconsistent with observations as described above.

Such an early break at 250~s, can perhaps be explained as the onset time of the
afterglow. If the reverse shock is non relativistic (as indicated by the
relatively short duration of the burst, see \citet{sari97}) then self
similar expansion starts once the mass collected from the environment is a
factor $\gamma$ smaller than that in the ejecta:
\begin{equation}
t_{{\rm afterglow}}=100\,{\rm s}\,(1+z) 
\left( \frac{E}{10^{53}\,{\rm erg}} \right)^{1/3} \left( \frac{n}{1\,{\rm
    cm^{-3}}} \right)
\left(\frac{\gamma}{100} \right)^{-8/3}.
\end{equation}
A value of the initial Lorentz factor $\gamma$ just below a hundred would
therefore be consistent with an onset time of 250~s. However, it is difficult
to reconcile the flat part before 250~s as the rise of the afterglow. During
the onset, since the fireball is coasting with a constant Lorentz factor, the
bolometric luminosity is given by $L_\mathrm{B} \propto t^2 n$, the surface
area times the density. For a constant density a sharp rise $\propto t^2$ is
therefore expected. For a wind-like decreasing density, the lightcurve should
be flat as observed. However, as stated before, a wind density profile seems
inconsistent with the behavior after 250~s.

Continuous energy injection has been suggested as a source of early X-ray
light\-curve flat\-tening~\citep{nkgpg05}. This injection could be observed if
the initial fireball ejecta had a range of Lorentz factors, with the slower
shells catching up with the decelerating afterglow~\citep{rm98,sm00}.  However,
we require a very steady injection of energy to produce the observed
lightcurve, flat for more than a decade in time.  If we adopt this explanation,
the afterglow must start before our first optical observation, implying an
initial Lorentz factor of more than 200, and energy injection rate which is
roughly constant over a decade in time, and which shuts off suddenly at 250~s.

Flat or very slowly decaying optical lightcurves have been seen in a
number of other early afterglows (eg, GRB~030418~\citep{rspaa04},
GRB~050319~\citep{qryaa05}, and GRB~041006~\citep{msmy04,ysr04}).  Early
X-ray lightcurves detected by \emph{Swift} are typically more complex,
with rapidly fading sections and short timescale flares~\citep{nkgpg05}.
The early afterglow of GRB~050801, flat in both optical and X-rays, is, 
so far, unique. It is inconsistent with the standard fireball model for
early afterglow emission, unless continuous energy injection is
involved. Further \emph{Swift} prompt GRB detections, combined with
rapid follow-up by \emph{Swift} and ground-based telescopes, will
provide further opportunities to explore the origin of this type of early
afterglow behavior.

\acknowledgements 

This work has been supported by NASA grants NNG-04WC41G and NGT5-135, NSF
 grants AST-0407061, the Australian Research Council, the University of New
 South Wales, and the University of Michigan.  Work performed at LANL is
 supported through internal LDRD funding. The Palermo work is supported at INAF
 by funding from ASI on grant number I/R/039/04.  Special thanks to Toni Hanke
 at the H.E.S.S. site.



\newcommand{\noopsort}[1]{} \newcommand{\printfirst}[2]{#1}
  \newcommand{\singleletter}[1]{#1} \newcommand{\switchargs}[2]{#2#1}




\end{document}